\documentclass[prl,aps,showpacs,twocolumn,superscriptaddress]{revtex4-1}
\usepackage{amsmath,amssymb,graphicx}
\usepackage{amsmath}

\begin{document}

\title{Scattering of a cross-polarized linear wave by a soliton at an optical event horizon in a birefringent nanophotonic waveguide}

\author{Charles Ciret}
\author{Simon-Pierre Gorza}
\email{charles.ciret@ulb.ac.be}
\affiliation{OPERA-Photonique, Universit\'e libre de Bruxelles (ULB), 50 av. F.D. Roosevelt, CP194/5, B-1050 Bruxelles, Belgium}




\begin{abstract}
The scattering of a linear wave on an optical event horizon, induced by a cross polarized soliton, is experimentally and numerically investigated in integrated structures. The experiments are performed in a dispersion-engineered birefringent silicon nanophotonic waveguide. In stark contrast with co-polarized waves, the large difference between the group velocity of the two cross-polarized waves enables a frequency conversion almost independent on the soliton wavelength. It is shown that the generated idler is only shifted by 10 nm around 1550 nm over a pump tuning range of 350 nm. Simulations using two coupled full vectorial nonlinear Schr\"odinger equations fully support the experimental results.
\end{abstract}


\maketitle

Nonlinear interactions between linear waves and solitons attract a tremendous amount of interest in the scientific community for many years. In this framework, interactions between waves of similar group velocities particularly draw the attention of researchers for their potential useful applications such as frequency converters \cite{Tartara:12:IEEE.J.Quantum.Electon.} or for future optical transistor-like devices \cite{Demircan:11:Phys.Rev.Lett}. In this particular process, the propagation of an intense solitonic pump in a Kerr medium induces a moving refractive index perturbation which in turn leads to a frequency conversion of a weak probe wave through a cross-phase modulation (XPM) process. In the context of supercontinuum (SC) generation, this process helps understanding the underlying physics of the spectral broadening \cite{Skryabin:10:RevModPhys}, and has also been demonstrated to enable the generation of highly coherent broadband SC \cite{Demircan:13:Phys.Rev.Lett,Demircan:14:optExpress} in a complementary manner from the well-known process involving the soliton fission \cite{Dudley:06:Rev.Mod.Phys,Leo:15:OL,Dave:15:OL}. Recently, this nonlinear interaction has been reinterpreted as the optical analog of the event horizon of black and white holes \cite{Philbin:08:Science,Webb:14:Nature.Commun}. As the propagation takes place in a dispersive media, the frequency conversion of the probe wave is accompanied by a modification of its group velocity, which is either accelerated or decelerated preventing any crossing between the two waves. The intense pulse thus constitutes an horizon that light can neither join nor escape.  These so-called optical event horizons have thus also been largely studied for their analogy with general relativity and in particular with the Hawking radiation \cite{Philbin:08:Science}.

Since its first theoretical prediction made by Yulin \textit{et al.} a decades ago \cite{Yulin:04:OL,Yulin:05:PRE}, and its experimental observation in the context of SC generation in optical fibers \cite{Efimov:05:PRL}, numerous experimental demonstrations have been realized. We can cite studies involving the superimposition of a linear wave to an intense pump at the waveguide input \cite{Philbin:08:Science,Webb:14:Nature.Commun,Ciret:16:OE}, or the interaction between two pulses \cite{Tartara:12:IEEE.J.Quantum.Electon.}. More recent studies have also investigated the interaction of an intense pulse with its own dispersive wave (DW) \cite{Wang:15:PRA} or the trapping of a DW between two solitons \cite{Wang:15:OL,Yulin:13:OptExpress}. Interactions involving higher order solitons \cite{Oreshnikov_1:15:OL} or dark solitons \cite{Oreshnikov_2:15:OL} have also been considered. Owing to the essential role played by the dispersion properties of the structure for the observation of an optical event horizon, almost all these demonstrations have been performed in photonic crystal fibers (PCF's), for which quasi on-demand dispersion properties can be engineered. Its observation in integrated structure has been reported only recently \cite{Ciret:16:OE}. The fully CMOS compatibility of the dispersion-engineered nanophotonic waveguide used in this latter demonstration as well as the absence of a broad Raman gain could be very useful for applications. 

 Almost all of the previous studies have focused on the interaction between waves having the same polarization. Orthogonally polarized pump and probe waves have only been considered in the context of birefringent PCF's \cite{Efimov:05:PRL}. In this paper, we report on the first experimental demonstration of the interaction that arises at the optical event horizon between two cross polarized waves propagating in an integrated fully CMOS compatible birefringent nanophotonic waveguide. The dispersion properties of the two cross polarized waves have been engineered through the waveguide dimensions, in order to allow for the two waves to interact. We show that the large difference between the group velocity dispersion of the two polarizations supported by our birefringent waveguide does not prevent the nonlinear interaction to occur, providing that the proper phase matching condition is met. However, this phase matching condition is only weakly dependent on the pump wavelength for the waveguide considered in this work. As a result the wavelength of the generated idler wave is almost independent on the soliton wavelength. This is clearly different from the previously reported co-polarized case. The ability to tailor the nonlinear interaction through the different dispersion properties of the signal and the pump waves opens new possibilities, similarly to the works performed in the context of the sum frequency generation in thick type II nonlinear crystals \cite{Wasylczyk:07:JModOpt}. The experimental results are fully supported by numerical simulations using two coupled full-vectorial generalized nonlinear Schr\"odinger equations. 

 \begin{figure}[b!]
\centering\includegraphics[width=\linewidth]{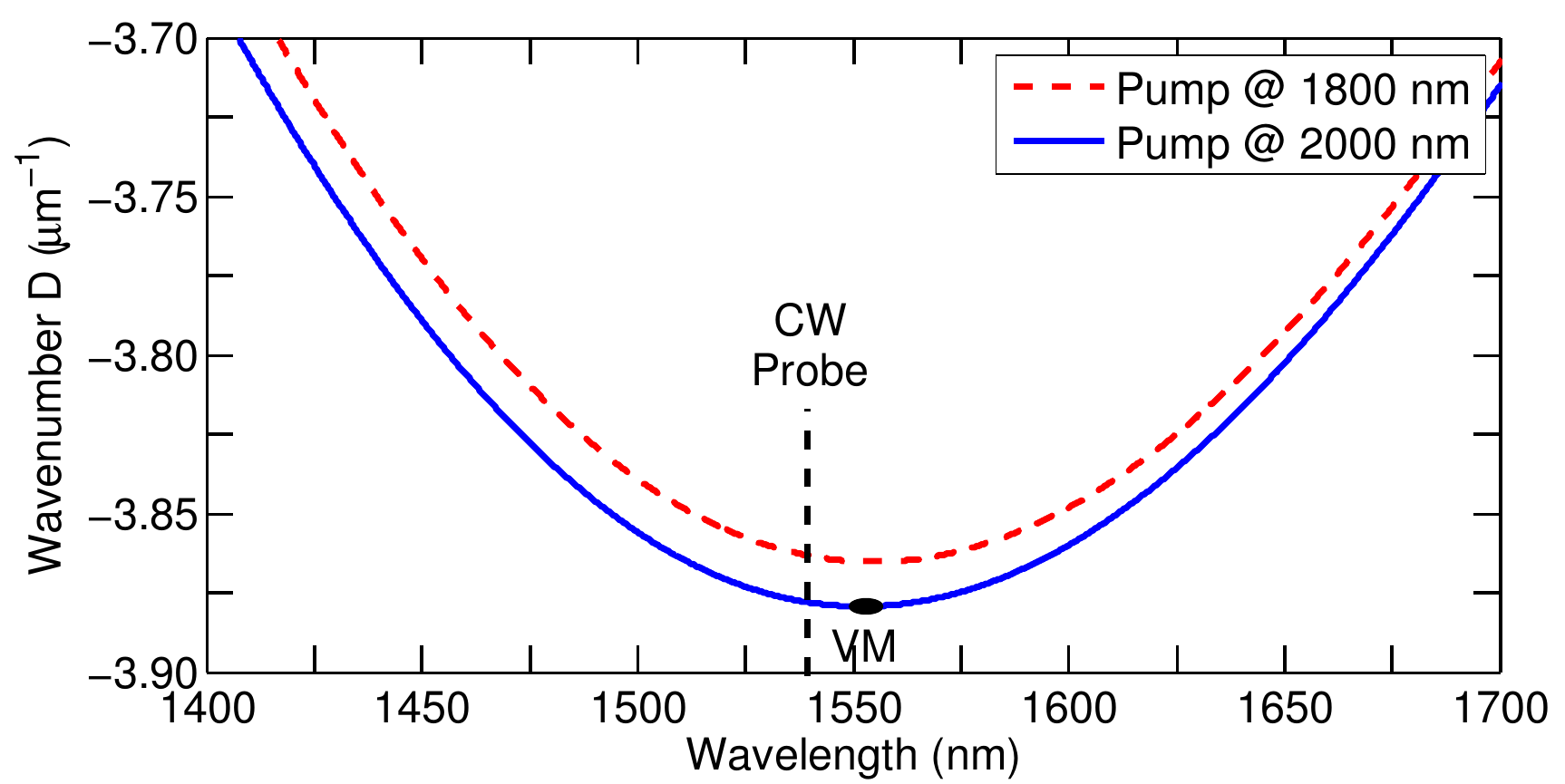}
\caption{Wavenumber $D$ as a function of the wavelength for two pump wavelengths (birefringent case, see in the text). The silicon nanophotonic waveguide under consideration is 220 nm-thick and 800 nm-wide. The vertical dashed line represents the position of a continuous probe wave (CW probe) at 1540 nm. VM point: group velocity-matched wavelength.}
\label{D}
\end{figure}

When interacting in a Kerr media, two waves with similar group velocities can lead to the generation of an idler wave set by the well-known resonant condition~\cite{Yulin:05:PRE,Efimov:05:PRL,Webb:14:Nature.Commun,Xu:13:OL}:

\begin{equation}
D(\omega_{idler}-\omega_{pump})=D(\omega_{probe}-\omega_{pump}).
\label{PMEq}
\end{equation} 
In this work, as two cross-polarized waves are considered, $D$ represents the wavenumber of a linear TM wave at a frequency $\omega$ in a reference frame co-moving with the TE pump at a frequency $\omega_{pump}$.  The wavenumber $D$ can be expressed as: $D(\omega-\omega_{pump})=\beta(\omega)_{_{TM}}-\beta_{0_{TE}}-\beta_{1_{TE}}\times(\omega-\omega_{pump}),$ where $\beta_{0_{TE}}=\beta(\omega_{pump})_{_{TE}}$ and $\beta_{1_{TE}}=d\beta_{_{TE}}/d\omega|_{\omega_{pump}}$. The wavelength dependence of the wavenumber $D$ for the waveguide used in the experiments is displayed in Fig.~$\ref{D}$  for two different (TE) pump wavelengths. In this figure the VM point (i.e. the zero-slope point) is the wavelength of the TM mode that is group velocity matched with the TE pump wave. Considering a particular probe wavelength, the wavelength of the generated idler wave is located on the other side of the VM point. The wavelength conversion thus prevents the probe wave to cross the pump pulse which constitutes its horizon.

Our demonstration is realized in a 6~mm-long birefringent silicon-on-insulator nanophotonic waveguide, with a standard height of 220~nm and a width of 800~nm. This waveguide enables the propagation of both the fundamental quasi-TE and quasi-TM modes at wavelengths of interest, as confirmed by finite difference simulations performed using MODE Solutions\copyright$ $ from Lumerical. Because of the subwavelength dimensions of the structure and the high index contrast between the guiding and the surrounding regions, the two fundamental modes have non-negligible electric and magnetic longitudinal components \cite{Afshar:09:OE}. These longitudinal components could lead to the coupling between the two polarizations during the propagation and modify the nonlinear coefficients usually encountered in the generalized scalar nonlinear Schr\"odinger equation. The set of two-coupled full vectorial nonlinear Schr\"odinger equations used to describe the propagation in our highly birefringent waveguide can be expressed as \cite{Afshar:09:OE,Daniel:10:JOSAB,Afshar:12:OE}:  
 
\begin{align}
\frac{\partial a_\nu}{\partial z}&=i\sum_{n=1}^\infty\frac{(i\partial/\partial t)^n}{n!}\beta^n_\nu a_\nu \nonumber -\frac{\alpha_\nu}{2}a_\nu\\&+(1+\tau_{shock}\partial/\partial t)\left[ i\gamma_\nu |a_\nu |^2 a_\nu + 2i\gamma_{\mu\nu}|a_{\mu}|^2a_\nu \right],
\label{VNLSE}
\end{align} 
in which the subscripts $\nu,\mu=1,2$, with $\nu\neq\mu$, and $a_1(z,t)$, $a_2(z,t)$, denote the amplitudes of the quasi-TE and quasi-TM modes respectively. In these equations, $\beta_\nu^n=\partial^k\beta_\nu/\partial \omega$ are the dispersion coefficients associated with the Taylor series expansion of the propagation constant of each modes $\beta_\nu(\omega)$ around the pump frequency $\omega_{pump}$. These terms are responsible for the linear dispersion effect. In the simulations, as recommended by \cite{Dudley:06:Rev.Mod.Phys}, the dispersion is however directly applied without any approximation in the frequency domain, from the function $\beta_\nu(\omega)$ evaluated for both polarizations using finite difference simulations. $\alpha_\nu$ are the linear loss coefficients. They are assumed to be constant at 2 dB/cm for both polarizations on the wavelength range of interest. The dispersion of the nonlinearity is taken into account by the shock terms, as commonly used in previous studies \cite{Dudley:06:Rev.Mod.Phys,Ciret:16:OE}. Finally, $\gamma_\nu, \gamma_{\mu\nu}$ are the complex nonlinear parameters which are responsible for the self- and the cross-phase modulation effects, respectively. They are given by:

\begin{align}
\gamma_{\nu}&=\frac{2\pi}{\lambda_\nu}\frac{\overline{n_{2_{\nu}}}}{A_{eff_{\nu}}}+i\frac{\overline{\beta_{TPA_{\nu}}}}{2A_{eff_{\nu}}} \nonumber\\
\gamma_{\mu\nu}&=\frac{2\pi}{\lambda_\nu}\frac{\overline{n_{2_{\mu\nu}}}}{A_{eff_{\mu\nu}}}+i\frac{\overline{\beta_{TPA_{\mu\nu}}}}{2A_{eff_{\mu\nu}}},
\label{gamma}
\end{align}
in which, $A_{eff_\nu}$ and $A_{eff_{\mu\nu}}$ are the generalized effective mode area and the overlapping generalized mode area between the two modes, respectively. $\overline{n_{2_{\nu}}}$, $\overline{n_{2_{\mu\nu}}}$ and $\overline{\beta_{TPA_{\nu}}}$, $\overline{\beta_{TPA_{\mu\nu}}}$ are the effective nonlinear refractive indexes and the effective two-photon absorption coefficients, averaged over an inhomogeneous cross section weighted with respect to the field distributions \cite{Afshar:09:OE,Daniel:10:JOSAB,Afshar:12:OE}. In Table \ref{tab} the value of the nonlinear parameters and of the generalized effective areas for both polarizations as well as the overlapping  generalized effective area are given at the pump (TE) and the CW probe (TM) wavelengths. They have been estimated by calculating the electric and magnetic components of the mode profiles with finite difference simulations and from the values of the nonlinear indexes and the two photon absorption coefficients reported in the literature \cite{Bristow:07:APL}. Note that in Eq.~(\ref{PMEq}), the free carriers dispersion and absorption as well as the Raman effect have been neglected, as they have a marginal impact on the nonlinear dynamics in our experimental conditions \cite{Ciret:16:OE}. This can simply be understood by the absence of a broad Raman gain close to the pump wavelength and by the low pulse energy and the carrier relaxation time ($\approx$ 1~ns) which is much shorter than the delay between two pulses (12~ns). As shown in \cite{Ciret:16:OE}, the absence of the Raman soliton self frequency shift is beneficial for potential applications of optical event horizons.

\begin{table}[t!]
\centering
\caption{\bf Values of the coefficients of Eq.~(\ref{gamma}), corresponding to the experiments.}
\begin{tabular}{ccc}
\hline
 &TM-mode & TE-mode \\
\hline
 $\lambda$ (nm) & 1540  & 2000  \\
$\gamma_\nu$ (W$^{-1}$m$^{-1}$) &  201+38i& 220+18i \\
$\gamma_{\mu\nu}$ (W$^{-1}$m$^{-1}$) & 66+7i& 70+12i \\
$A_{eff_\nu}$ ($\mu$m$^{2}$) & 0.324 & 0.205 \\
$A_{eff_{\mu\nu}}$ ($\mu$m$^{2}$) & 0.287& 0.287 \\
\hline
\end{tabular}
  \label{tab}
\end{table}

\begin{figure}[b!]
\centering\includegraphics[width=\linewidth]{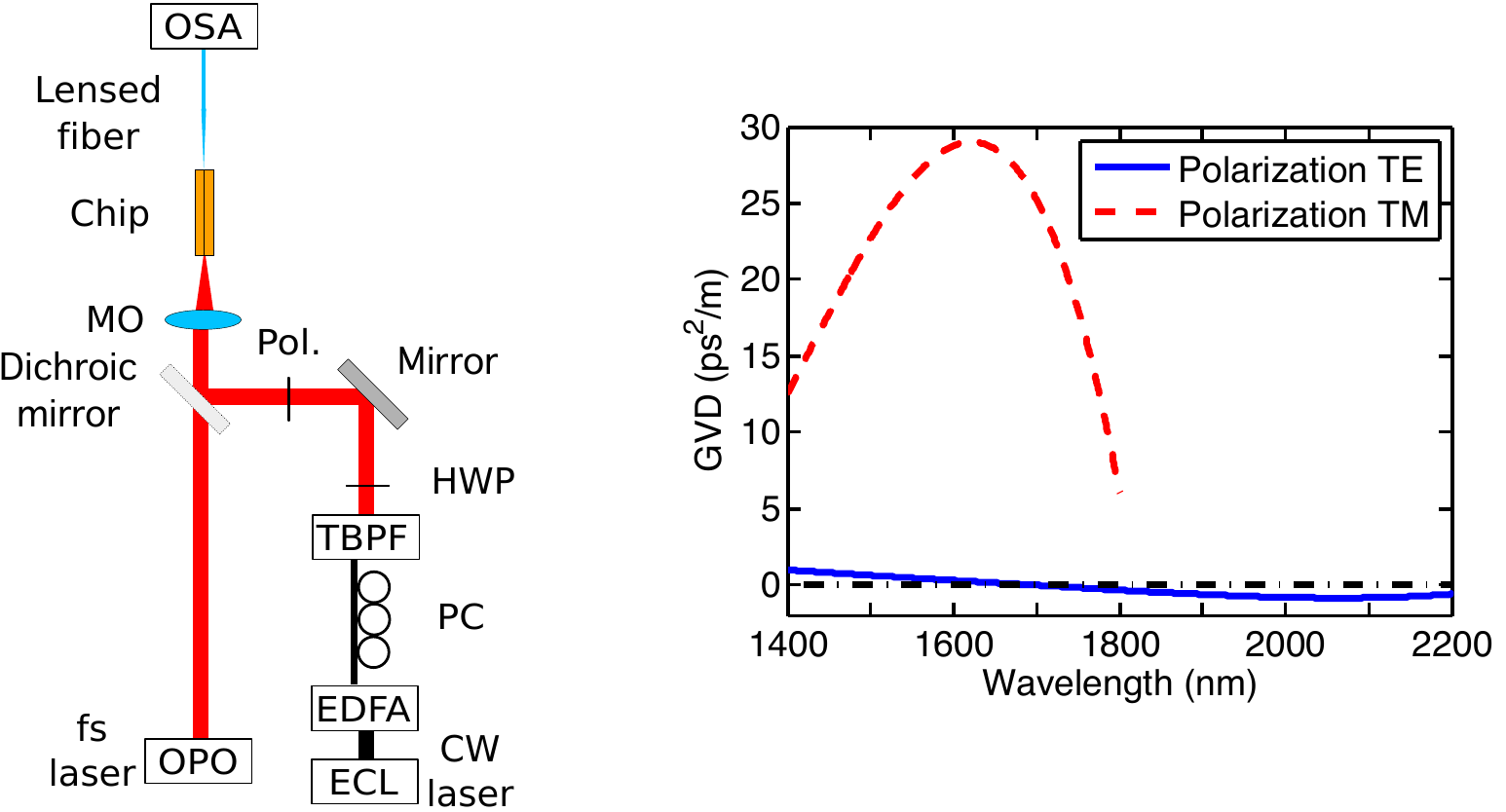}
\caption{Left: Experimental setup. OPO: optical parametric oscillator, ECL: external cavity laser, EDFA: erbium doped fiber amplifier, PC: polarization controller, TBPF: tunable bandpass filter, HWP: half wave-plate, Pol: polarizer, MO: microscope objective, OSA: optical spectrum analyzer. Right: Simulated group velocity dispersion curves for the TE-polarized (blue line curve) and the TM-polarized (red dashed curve) fundamental mode of the waveguide used in the experiments. }
\label{setup_GVD}
\end{figure}

\begin{figure}[b!]
\centering\includegraphics[width=\linewidth]{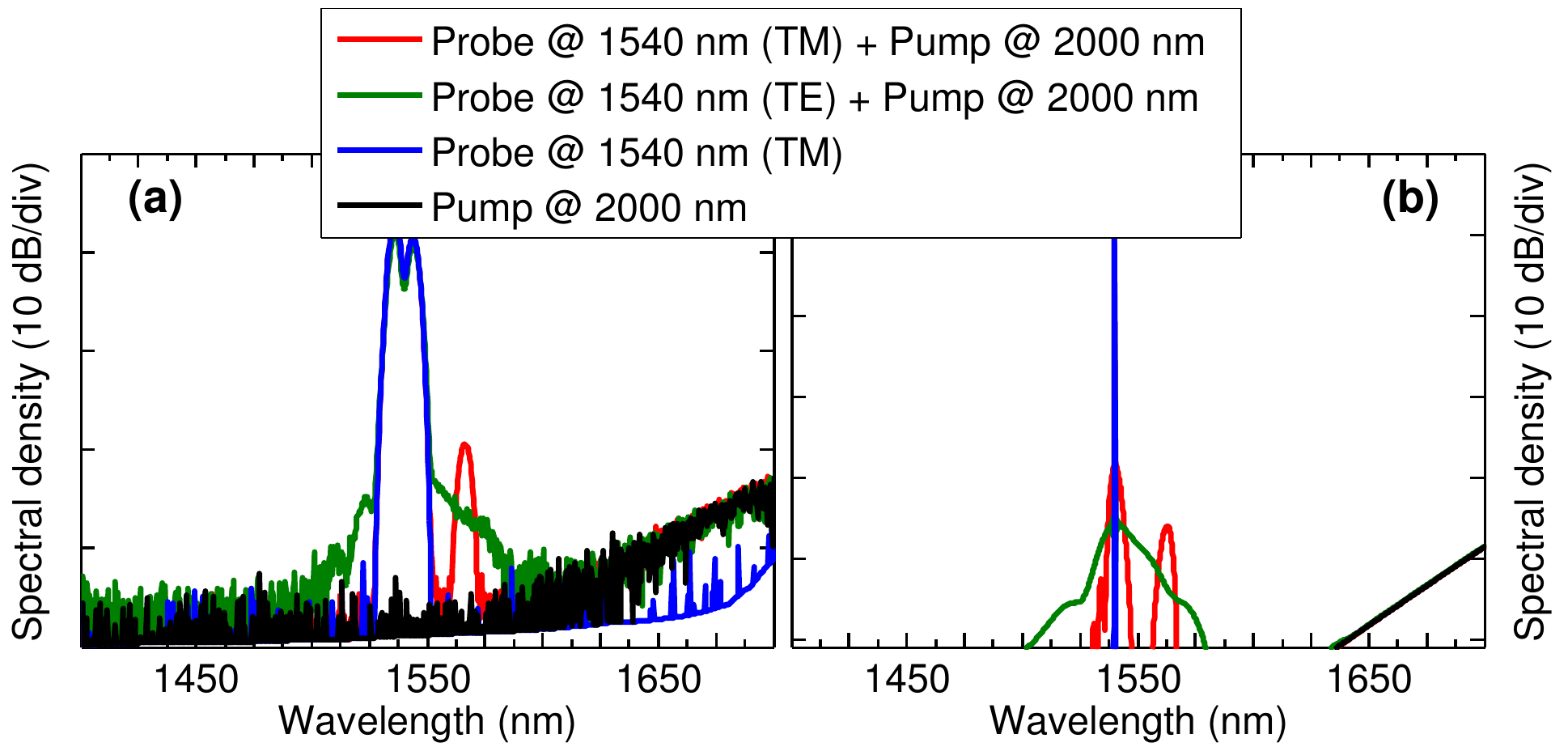}
\caption{(color online) (a): Experimental spectra recorded, with a 5~nm resolution, at the output of the 6~mm-long Si waveguide and (b): the corresponding simulated spectra (b). }
\label{EventH-ExpSimu}
\end{figure}

The experimental setup and the simulated dispersion curves for both polarizations are shown  in Fig.~\ref{setup_GVD}. The nanowire possesses a low dispersion for the quasi-TE mode with a zero dispersion wavelength at 1700~nm, whereas for the TM mode the dispersion is always high and normal in the range 1400~nm-1800~nm. Beyond this wavelength, the TM mode experiences a high linear loss. The input pump pulse  is generated by an optical parametric oscillator delivering 180~fs pulses (full width at half maximum) at a wavelength of 2000~nm and with a 82 MHz repetition rate. Its polarization excites the fundamental quasi-TE mode of the structure. The CW probe is generated by a CW external cavity laser followed by an Er-doped fiber amplifier. Its wavelength can be tuned from 1525~nm to 1575~nm. At the amplifier output, the broad amplified spontaneous emission is filtered out by a tunable bandpass filter. The two collimated beams are combined on a dichroic mirror and coupled into the waveguide by means of a microscope objective. The polarization of the CW probe can be carefully adjusted using a half wave-plate and a Glan-Thompson polarizer in order to excite either the fundamental quasi-TE or quasi-TM mode of the nanowire. At the waveguide output, the light is collected by a lensed fiber, then passes through a narrow filter to attenuate the remaining CW probe for a better visualization of the spectral features close to that wavelength. The filtered spectrum is recorded by an optical spectrum analyzer sensitive in the range 700~nm-1700~nm. The experimental output spectra are displayed in Fig.~\ref{EventH-ExpSimu}(a) for an on-chip peak pump power of 3~W and a 500~$\mu$W CW probe at 1540~nm. At such peak power, the propagation of the pump alone in the anomalous dispersion region leads to the generation of a fundamental soliton. Only part of its pedestal can be seen in the spectrum as the black curve in Fig.~\ref{EventH-ExpSimu}(a). The spectrum of the TM-polarized CW probe alone is shown as the blue curve. When both the TM CW probe and the TE pump are propagating together, we can clearly observe the emergence of an idler wave at 1565~nm. This wavelength is in very good agreement with the phase matching condition Eq.~(\ref{D})  represented as the blue curve in Fig.~\ref{setup_GVD} for a pump wavelength of 2000~nm. When the polarization of the CW probe is rotated from TM to TE, only a wide pedestal close to the probe wavelength can be observed. This broadening is the result of the cross-phase modulation induced on the CW probe by the soliton pulse. Further propagation shows the emergence of an idler wave as demonstrated in \cite{Ciret:16:OE}. The simulations performed by solving the Eqs.~(\ref{VNLSE}) are plotted in Fig.~\ref{EventH-ExpSimu}(b). These results are in excellent agreement with the recorded experimental spectra. This confirms both the validity of the vectorial generalized nonlinear Schr\"odinger model Eq.~(\ref{VNLSE}) and the interpretation of the experimental results. The higher efficiency of the frequency conversion when two cross-polarized waves are interacting can be explained by the dispersion properties of the two modes. As shown in~\cite{Tartara:12:IEEE.J.Quantum.Electon.}, the conversion efficiency for a CW probe scales linearly with the group velocity difference between the probe and the soliton and, in our waveguide, this difference is five times larger at 1540~nm for the TM probe than for the TE one.

\begin{figure}[b!]
\centering\includegraphics[width=\linewidth]{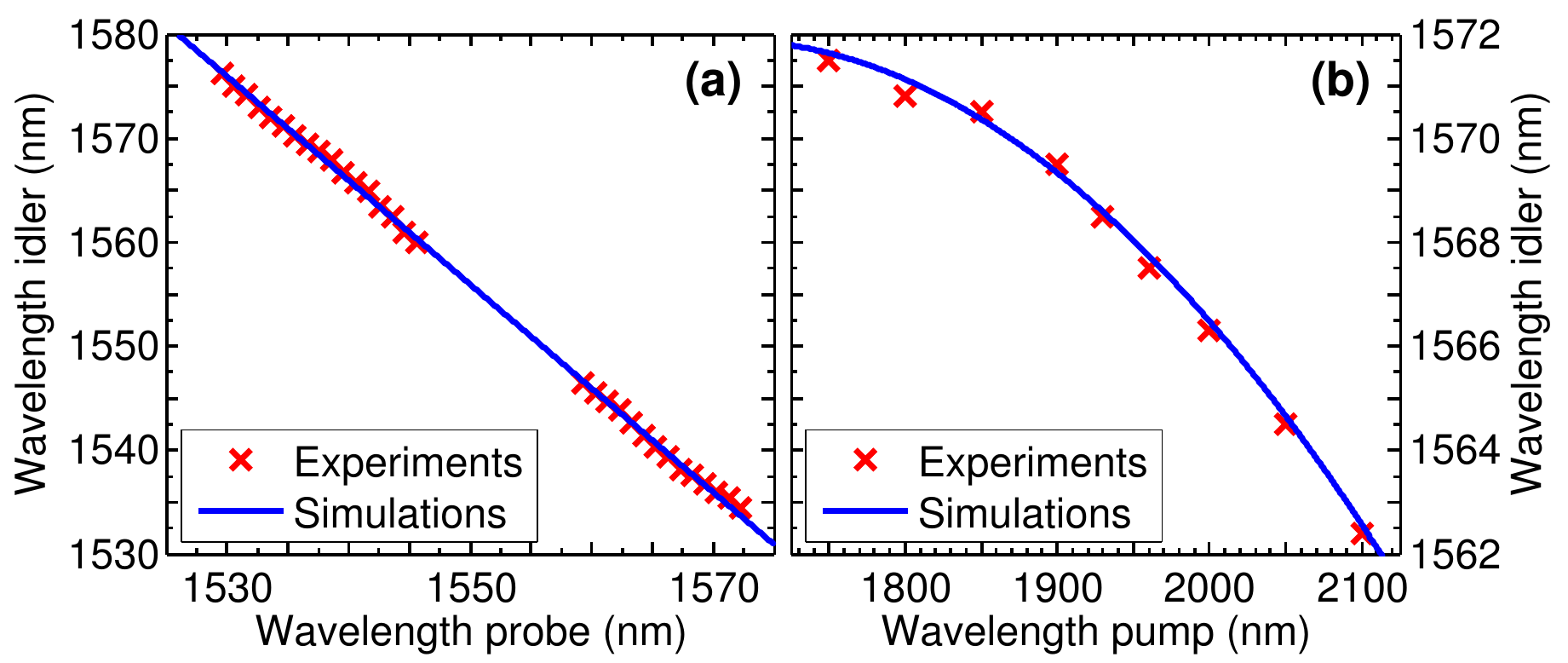}
\caption{(a): Wavelength of the idler wave as a function of the CW probe wavelength for a pump at 2000 nm. There are no experimental results in the range 1545~nm-1555~nm because the wavelengths of the CW probe and the idler are too close to be distinguishable. (b): Evolution of the idler wavelength with the pump wavelength for a CW probe at 1540 nm. Note that the results from  Eq.~(\ref{PMEq}) are not shown as they are exactly the same as the simulation curves.}
\label{Tuning}
\end{figure}


The probe wave reflection on the optical event horizon is accompanied by a shift of its frequency leading to the appearance of the idler wave. This wavelength shift can be predicted on the basis of the phase matching condition Eq.~(\ref{PMEq}) generalized for a cross polarized interaction. It can simply be visualized in Fig.~\ref{D} as a reflection around the VM point. The wavelength dependence of the idler wave as a function of the pump wavelength is plotted in Fig.~\ref{Tuning}(a). The experimental results are in very good agreement with the simulations and with Eq.~(\ref{PMEq}). As can be seen, as the probe wavelength is tuned closer to the VM point, the idler wavelength also gets closer to that particular wavelength. More interestingly, it is shown in Fig.~\ref{Tuning}(b) that for a fixed probe wavelength of 1540~nm, the idler wavelength is almost independent of the soliton wavelength. As the soliton wavelength is tuned from 1750~nm up to 2100~nm, the idler wavelength is only shifted by 10~nm. This can be explained from the very low group velocity dispersion coefficient of the quasi-TE mode compared to the quasi-TM mode. As a result of this large difference, the particular wavelength corresponding to the VM point around which the reflexion of the probe on the optical event horizon occurs is only slightly shifted over a large pump wavelength range (see in Fig.~\ref{D}). This behavior is not encountered for co-polarized TE interactions as the dispersion coefficients are closer for the signal and the pump waves (see in Fig.~\ref{setup_GVD}). This interesting feature could for instance be useful for the realization of a broadband wavelength detector of pulses around 2000~nm, as the idler wave emerge in the C-band where sensitive optical detectors are available. Finally, note that no spectral recoil of the pump, associated with the scattering of the probe, has been observed in the experimental or simulated spectra. This comes from the too low energy transfer, inherent to the scattering of the low CW wave, as shown in \citep{Ciret:16:OE} for co-polarized interactions. The simulations with a pulsed probe reveal however that in this latter case a spectral recoil of the pump is observable and is associated with a 70~$\%$ energy conversion of the probe (not shown).     

In conclusion we have experimentally and numerically investigated the scattering of a cross polarized linear wave by a soliton at an optical event horizon in a birefringent nanophotonic waveguide. The simulations where performed by integrating a set of two coupled vectorial generalized nonlinear Schr\"odinger equations. In these equations, the longitudinal component of the electric and magnetic fields have been taken into account owing to the high index contrast subwavelength nanophotonic waveguides considered in this work. These components are responsible for the coupling between the two cross polarizations for low birefringent waveguides and change the expressions of the nonlinear coefficients. As the birefringence of our dispersion engineered waveguide is large, the longitudinal components have only been taken into account in the expression of the nonlinear coefficients. In the experiment, we have unambiguously observed the emergence of an idler wave when an intense quasi-TE pump pulse propagates together with a quasi-TM CW probe. The large difference in the group velocity dispersion coefficients for the two polarizations leads to a high net conversion efficiency as well as a low dependence with the pump wavelength on the generated idler wavelength, located in the C-band in this work. These results, very well supported by the numerical simulations, are in contrast with previously reported results and could be useful for potential integrated applications.

\section*{Acknowledgements}

This work is supported by the Belgian Science Policy Office (BELSPO) Interuniversity Attraction Pole (IAP) project Photonics@be and by the Fonds de la Recherche Fondamentale Collective, Grant No. PDR.T.1084.15. The authors thank B. Kuyken, G. Roelkens, U. D. Dave and F. Leo for the fruitful discussions and for providing the Si-waveguide.




\end{document}